\documentclass[reprint,amsmath,amssymb, aps, prb]{revtex4-2}

\usepackage{graphicx}
\usepackage{bm}
\usepackage{color}

\begin{document}
\title{
  Tensor-network strong-disorder renormalization groups for random quantum spin systems in two dimensions
}

\author{Kouichi Seki$^1$}
\author{Toshiya Hikihara$^2$}
\author{Kouichi Okunishi$^1$}

\affiliation{
  $^1$Department of Physics, Niigata University, Niigata 950-2181, Japan \\
  $^2$Faculty of Science and Technology, Gunma University, Kiryu, Gunma 376-8515, Japan
}

\begin{abstract}
Novel randomness-induced disordered ground states in two-dimensional (2D) quantum spin systems have been attracting much interest.
For quantitative analysis of such random quantum spin systems, one of the most promising numerical approaches is the tensor-network strong-disorder renormalization group (tSDRG), which was basically established for one-dimensional (1D) systems. 
In this paper, we propose a possible improvement of its algorithm toward 2D random spin systems, focusing on a generating process of the tree network structure of tensors, and precisely examine their performances for the random antiferromagnetic Heisenberg model not only on the 1D chain but also on the square- and triangular-lattices.
On the basis of comparison with the exact numerical results up to 36 site systems, we demonstrate that accuracy of the optimal tSDRG algorithm is significantly improved even for the 2D systems in the strong-randomness regime.
\end{abstract}

\date{\today}

\maketitle

\section{Introduction}
Quenched randomness is a source of intriguing phenomena in quantum spin systems.
For example, the ground state of the one-dimensional (1D) antiferromagnetic (AF) Heisenberg model with exchange randomness is well-described by the random singlet state, which is a product state of spin-singlet pairs distributed according to the spatial profile of the exchange-coupling constants.\cite{MaDH1979,DasguptaM1980,Fisher1994,YusufY2003}
For the two-dimensional (2D) random AF Heisenberg models with frustration, it was suggested that the randomness and frustration cooperatively induced a novel nonmagnetic disordered ground state with peculiar thermodynamic properties,\cite{WatanabeKNS2014, ShimokawaWK2015, WuGS2019, KawamuraWS2014, UematsuK2017, UematsuK2018, KawamuraU2019, RenXWSG2020}
which might relevant to spin-liquid-like behaviors observed in several quantum magnets.\cite{ShimizuMKMS2003,ItouOMTK2008,IsonoKUTNKNKMM2013,note_exp_ref}
Similar spin-liquid-like ground states were also suggested for the pyrochlore-lattice random AF Heisenberg model \cite{UematsuK2019} and the $J$-$Q$ model with randomness on the square lattice.\cite{LiuSLGS2018}
For the square-lattice random AF Heisenberg model without frustration, meanwhile, a small but finite N\'{e}el order survives in the bulk limit, although its amplitude is strongly suppressed by the randomness in exchange couplings.\cite{LaflorencieWLR2006}

Despite of the increasing interest in the 2D random quantum spin systems, theoretical studies of them have been challenging since analytical methods developed for homogeneous systems often lose their validity.
Numerical methods are relatively robust against randomness, but they also suffer from limitations; 
The exact diagonalization (ED) can treat only small systems, and the quantum Monte Carlo (QMC) method is not basically applicable to frustrated systems due to the minus-sign problem.
Thus, a numerical approach efficient in treating 2D quantum spin systems with randomness is highly desired.

For this purpose, we focus on the tensor-network strong-disorder renormalization group (tSDRG)\cite{HikiharaFS1999}.
The tSDRG was originally introduced as a numerical renormalization group for the 1D AF Heisenberg model with exchange randomness, inspired by the perturbative strong-disorder renormalization group (SDRG) approach.\cite{MaDH1979,DasguptaM1980} 
The tSDRG has then been successfully applied to a wide class of 1D random spin systems.\cite{GoldsbroughR2014,LinKCL2017,TsaiCL2019}
However, the efficiency of the tSDRG has not been sufficiently examined for 2D systems so far, and as we will discuss below, there is still room for improvement in its algorithm.

In the tSDRG, the system is represented as an assembly of blocks of original spins.
Then, the main process of the tSDRG algorithm is composed of the following two steps;
The first step is to find out the block pair connected by the ``strongest" link from any interacting pairs of blocks and the second one is to renormalize the pair into a new block with truncated bases.
The effective dimension of the Hilbert space of the whole system is thereby reduced, as the tSDRG iterations proceed.
The above process is continued until the whole system is represented by a few blocks such that the Hamiltonian of the whole system can be exactly diagonalized within the renormalized basis.
In this paper, we discuss an improvement of the first step of the tSDRG process to determine the strongest link, since this step directly reflects on the resulting network structure of tensors generally important for the numerical accuracy of tensor network algorithms.
In the previous tSDRG algorithm developed in Ref.\ \onlinecite{HikiharaFS1999}, a certain energy gap of the local Hamiltonian of block pairs was employed as a measure of the strength of the link.
We will propose tSDRG algorithms with several types of energy gaps that can be used as a measure of the strongest link and examine their performance.

In the following, we consider the $S=1/2$ AF Heisenberg model with the random nearest-neighbor interaction as a typical example of random quantum spin systems.
The Hamiltonian is formally written as,
\begin{equation}
  \mathcal{H}=\sum_{i, j} J_{i, j} \bm{S}_i \cdot \bm{S}_j,
\label{Hamiltonian}
\end{equation}
where $\bm{S}_i$ is a $S=1/2$ spin operator at $i$th site
and $J_{i, j}$ denotes the exchange-coupling constant.
For the nearest-neighbor coupling constants, i.e., $J_{i, j}$ with $(i,j)$ indicating the nearest-neighbor-site pairs, we assume the box-type randomness,
\begin{equation}
  P(J_{i,j}) =
    \frac{1}{2\delta} \Theta(J_{i,j}-1+\delta)\Theta(1+\delta-J_{i,j})
\label{eq_dist}
\end{equation}
where $\delta$ ($0 \le \delta \le 1$) is a parameter controlling the strength of the randomness
and $\Theta$ is the Heavyside step function.
We also set 
\begin{eqnarray}
J_{i,j}=0,
\label{eq_dist_nonNN}
\end{eqnarray}
if $(i,j)$ are not on the nearest neighbor sites.
We apply tSDRG algorithms with several definitions of the energy gap to Hamiltonian (\ref{Hamiltonian}) with Eqs. (\ref{eq_dist}) and (\ref{eq_dist_nonNN}) not only on 1D chain but also on the square and triangular lattices. 
Then, we demonstrate that the algorithm with the optimal choice of the gap actually improves the numerical accuracy of calculations
compared with the previous algorithm of Ref. \onlinecite{HikiharaFS1999}.

The paper is organized as follows.
In Sec.~\ref{sec:algorithm}, we review the tSDRG algorithm and introduce several energy gaps used for determining the strongest link connecting two blocks to be renormalized.
In Sec.~\ref{sec:numerics}, we present our results of tSDRG calculations and discuss their performances for the 1D-, square-, and triangular-lattice systems.
Sec.~\ref{sec:Summary} is devoted to summary and discussions.

\section{Algorithm}\label{sec:algorithm}
In this section, we discuss algorithms of the tSDRG in details.
The tSDRG was developed\cite{HikiharaFS1999,GoldsbroughR2014} as a numerical extension of the perturbative SDRG\cite{MaDH1979,DasguptaM1980}.
In order to grasp  possible improvements of tSDRG, thus, it is instructive to briefly review theoretical backgrounds of the SDRG approaches.

The perturbative SDRG was originally proposed as an analytic real-space renormalization group for the random AF Heisenberg spin chain.
The spin pair connected with the strongest exchange coupling is decimated to form the spin singlet and then the effective coupling between the two spins across the decimated singlet is evaluated with the second-order perturbation, which leads the analytic recursion relation for the distribution function of the random exchange coupling.
For the random AF chain, this recursion relation becomes asymptotically exact toward the random singlet fixed point in the bulk limit.\cite{Fisher1994}
Recently, the perturbative SDRG has been also extended to a variety of random spin systems such as  random Heisenberg chain with ferromagnetic and AF couplings\cite{WesterbergFSL1995,WesterbergFSL1997}, 2D random AF Heisenberg models,\cite{LinMRI2003} and transverse-field Ising models\cite{Fisher1992,Fisher1995,MotrunichMHF2000,LinIR2007}, for investigating the infinite-randomness fixed points.
However, we should note that the perturbative SDRG is not appropriate for quantitative calculations of physical quantities such as ground-state energy and correlation functions.
This is mainly because the perturbative approximation completely neglects contributions of local excited states to the ground state of the whole system in the early stage of its recursive calculation.

\begin{figure}[tb]
  \centering\includegraphics[width=\linewidth]{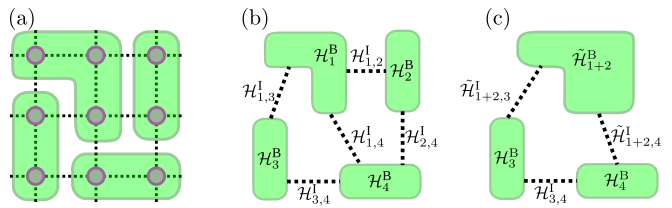}
  \caption{
    Schematic diagrams of the renormalization procedure of the tSDRG.
(a) The system after a certain number of tSDRG iterations, composed of blocks (green rectangles) of original spins (red circles).
(b) The effective Hamiltonian [Eq.\ (\ref{sq_SDRG_H})] consisting of the intra-block Hamiltonians $\mathcal{H}^{\rm B}_r$ and inter-block Hamiltonians  $\mathcal{H}^{\rm I}_{r,r'}$.
(c) The effective Hamiltonian after renormalizing the blocks of 1 and 2 into a new block ``1+2".
  }\label{fig_SDRG_1}
\end{figure}

In order to overcome the difficulty above, the tSDRG algorithm employs the block-state representation for the renormalized spin pair of the strongest coupling, retaining not only the lowest multiplet state but also excited states.
Suppose that after a certain number of tSDRG iterations,  spins on a regular lattice are merged into blocks as shown in Fig. \ref{fig_SDRG_1}(a).
The effective Hamiltonian of the whole system at this stage is written as
\begin{equation}
\mathcal{H} = \sum_r \mathcal{H}^{\rm B}_r + \sum_{r, r'} \mathcal{H}^{\rm I}_{r,r'},
\label{sq_SDRG_H}
\end{equation}
where $\mathcal{H}^{\rm B}_{r}$ is the renormalized Hamiltonian of $r$th block originating from the intra-block interactions,
\begin{eqnarray}
\mathcal{H}^{\rm B}_r = \sum_{i,j \in r} J_{i,j} {\bm S}_i \cdot {\bm S}_j,
\label{eq:Ham_intra}
\end{eqnarray}
and $\mathcal{H}^{\rm I}_{r, r'}$ represents the effective inter-block interaction between the $r$th and $r'$th blocks,
\begin{eqnarray}
\mathcal{H}^{\rm I}_{r,r'} = \sum_{i\in r, j\in r'} J_{i,j} 
{\bm S}_i \cdot {\bm S}_j.
\label{eq:Ham_inter}
\end{eqnarray}
[See Fig. \ref{fig_SDRG_1}(b).]
The dimension of the Hilbert space for each block is assumed to be truncated by $\chi$, so that the block Hamiltonian $\mathcal{H}^{\rm B}_{r}$ and the original spin operators $S^\alpha_i$ are represented as $\chi \times \chi$ matrices while the inter-block Hamiltonians $\mathcal{H}^{\rm I}_{r, r'}$ are $\chi^2 \times \chi^2$ matrices.
(To be precise, the dimension of the block basis is less than or equal to $\chi$, see below.)
Note that $\mathcal{H}^{\rm I}_{r, r'}$ is zero if the $r$th and $r'$th blocks are not connected via nonzero $J_{i,j}$.

In the tSDRG, two blocks connected by the ``strongest link", which is determined by a certain criterion discussed below, are merged into a new block.
Let $R$th and $R'$th blocks be the ones to be merged.
The Hamiltonian for the block pair is given by
\begin{eqnarray}
\mathcal{H}^{\rm P}_{R+R'} = \mathcal{H}^{\rm B}_{R} + \mathcal{H}^{\rm B}_{R'} + \mathcal{H}^{\rm I}_{R,R'}.
\label{eq:Ham_new_block}
\end{eqnarray}
Here, the dimension of the Hilbert space of the block pair is $\chi^2$, and one must truncate the space in order to avoid the exponential growth of its dimension.\cite{note_early_stage}
Using the $\chi$-lowest-energy eigenstates of the block-pair Hamiltonian (\ref{eq:Ham_new_block}), then, we renormalize the Hamiltonian of the new block as
\begin{eqnarray}
\tilde{\mathcal{H}}^{\rm B}_{R+R'} = V^{\dagger} \mathcal{H}^{\rm P}_{R+R'} V,
\label{eq:Ham_new_block_renorm}
\end{eqnarray}
where the renormalization matrix $V$ is composed of the $\chi$-lowest-energy eigenvectors $\{ {\bm v}_1, ..., {\bm v}_\chi\}$ of $\mathcal{H}^{\rm P}_{R+R'}$.
In practice, one must keep or discard all the eigenstates belonging to the same SU(2) multiplet in order to maintain the symmetry of the system, resulting that the dimension of the new renormalized block bases is $\chi' \le \chi$.
The matrices of the original spin operators belonging to the new block are also transformed as 
\begin{eqnarray}
\tilde{S}^\alpha_r &=& V^{\dagger} [S^\alpha_r \otimes I_{R'}] V~~~(r \in R),
\\
\tilde{S}^\alpha_{r'} &=& V^{\dagger} [I_{R} \otimes S^\alpha_{r'}] V~~~(r' \in R'),
\end{eqnarray}
where $I_{R'}$ ($I_R$) is the identity matrix for the block $R'$ ($R$).
Likewise, the inter-block Hamiltonians between the new block $R+R'$ and a block $R''$ connected to the new block by nonzero $\mathcal{H}^{\rm I}_{R,R''}$ and $\mathcal{H}^{\rm I}_{R',R''}$ are transformed as
\begin{eqnarray}
\tilde{\mathcal{H}}^{\rm I}_{(R+R'),R''} = V^{\dagger} \left[ \mathcal{H}^{\rm I}_{R,R''}\otimes I_{R'} + \mathcal{H}^{\rm I}_{R',R''}\otimes I_{R} \right] V.
\nonumber \\
\end{eqnarray}

\begin{figure}[tb]
  \centering\includegraphics[width=0.8\linewidth]{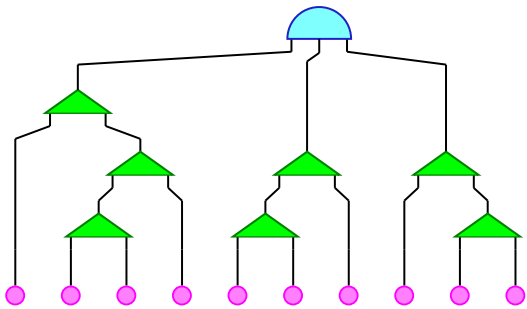}
  \caption{
    Schematic tree-tensor-network representation of the ground-state wavefunction obtained by the tSDRG.
    Circles (red) represent the original spins, and triangles (green) represent the renormalization-group transformation matrices.
    Semicircle (blue) at the top of the network represents the ground-state eigenvector of the effective Hamiltonian (\ref{sq_SDRG_H}) at the final step of the tSDRG.
  }\label{fig_tree}
\end{figure}

As the above tSDRG iteration recursively proceeds, the number of blocks in the system reduces one by one, and  the effective dimension of the Hilbert space of the whole system accordingly reduces.
In our calculation, we stop the tSDRG iteration when the number of blocks reduces down to three, where the effective Hamiltonian (\ref{sq_SDRG_H}) for the whole system can be exactly diagonalized within the truncated basis. 
Then, the resulting ground-state wavefunction can be represented as a tree tensor network schematically depicted in Fig. \ref{fig_tree}, for which one can straightforwardly compute expectation values of physical quantities such as energy and correlation functions.

An important point of Eq.\ (\ref{eq:Ham_new_block_renorm}) is that the renormalization process of $\tilde{\mathcal{H}}^{\rm B}_{R+R'}$ involves higher energy multiplets in addition to the lowest-energy multiplet, implying that the tSDRG is capable of nontrivial correlation effects neglected in the perturbative SDRG. 
This  point is a clear advantage of the tSDRG over the perturbative SDRG.
On the other hand, it also implies that information of the ground state of the total system is embedded in complex spectrum of iteration-number dependent block Hamiltonians, in contrast to the perturbative SDRG where the effective couplings between spins can be explicitly obtained with the perturbation theory.
In the tSDRG, thus, how to find out the ``strongest link" connecting the block pair turns out to be a rather nontrivial problem.

\begin{figure}[tb]
  \centering\includegraphics[width=0.95\linewidth]{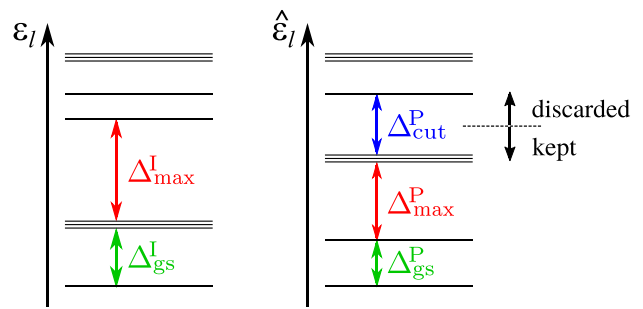}
  \caption{
    Schematic diagram of the energy gaps 
    $\{\Delta^{\! \rm I}_{\rm max}, \Delta^{\! \rm I}_{\rm gs}, \Delta^{\! \rm P}_{\rm max}, 
    \Delta^{\! \rm P}_{\rm gs}$, $\Delta^{\! \rm P}_{\rm cut} \}$ 
    introduced in the tSDRG algorithms.
    Eigenenergy spectrum $\{ \varepsilon_l \}$ of the inter-block Hamiltonian $\mathcal{H}^{\rm I}_{r, r'}$ is represented as horizontal lines in the left panel.
    The spectrum $\{ \hat{\varepsilon}_l \}$ of the block-pair Hamiltonian $\mathcal{H}^{\rm P}_{r+r'} = \mathcal{H}^{\rm B}_{r} 
    + \mathcal{H}^{\rm B}_{r'} + \mathcal{H}^{\rm I}_{r,r'}$ is also depicted in the right panel.
  }\label{fig_SDRG_scale}
\end{figure}

For the determination process of the strongest link, the tSDRG algorithm employs an energy gap in the spectrum of local Hamiltonians of block pairs as a measure of the strength of the links.
Here, there are several options and we propose the following ones.
Let $\{ \varepsilon_l \}$ ($1 \le l \le \chi^2$) be the eigenenergies of the inter-block Hamiltonian $\mathcal{H}^{\rm I}_{r, r'}$ arranged in ascending order.
From this energy spectrum, we define the following two energy gaps.
\begin{itemize}
\item The maximum energy gap in the spectrum, $\Delta^{\! \rm I}_{\rm max} \equiv  {\rm max} \{\varepsilon_{l+1} -\varepsilon_l\} $, which may capture the most significant physical structure embedded in $\mathcal{H}^{\rm I}_{r,r'}$.
\item The energy gap just above the ground state, $\Delta^{\! \rm I}_{\rm gs} \equiv \varepsilon_2 - \varepsilon_1$.
Here, if the ground states are degenerate, $\varepsilon_2$ should be replaced with that of the first excited state above the ground-state multiplet.
$\Delta^{\! \rm I}_{\rm gs}$ may capture the physical structure associated with the local ground state of $\mathcal{H}^{\rm I}_{r,r'}$.
\end{itemize}
In the same way, we can define two energy gaps $\Delta^{\! \rm P}_{\rm max}$ and $\Delta^{\! \rm P}_{\rm gs}$  also for the energy spectrum $\{ \hat{\varepsilon}_l \}$ of the block-pair Hamiltonian $\mathcal{H}^{\rm P}_{r+r'} = \mathcal{H}^{\rm B}_{r} + \mathcal{H}^{\rm B}_{r'} + \mathcal{H}^{\rm I}_{r,r'}$.
In addition, we consider the following energy gap:
\begin{itemize}
\item The energy gap between the highest energy in the states to be kept and the lowest energy in the states to be discarded, $\Delta^{\! \rm P}_{\rm cut} \equiv \hat{\varepsilon}_{\chi+1} - \hat{\varepsilon}_\chi$.
If the $\chi$th lowest-energy state is degenerate, $\chi$ is replaced with $\chi'$($\le \chi$) to maintain the SU(2) symmetry as mentioned above.\cite{note_Delta_cut}
\end{itemize}
Note that $\Delta^{\! \rm P}_{\rm cut}$ is equivalent to the gap employed in the previous tSDRG\cite{HikiharaFS1999}, which certainly works for the random AF Heisenberg chain. 
Figure\ \ref{fig_SDRG_scale} schematically illustrates the gaps introduced.

Let $\Delta$ be one of the above five gaps: $\Delta \in \{ \Delta^{\! \rm I}_{\rm max}, \Delta^{\! \rm I}_{\rm gs}, \Delta^{\! \rm P}_{\rm max}, \Delta^{\! \rm P}_{\rm gs}, \Delta^{\! \rm P}_{\rm cut} \}$.
Given definition of $\Delta$, we extract the spin pair $(R, R')$ having the maximum $\Delta$ from the all block pairs $(r, r')$ connected with  nonzero $\mathcal{H}^{\rm I}_{r,r'}$ and then perform the renormalization transformation of Eq. (\ref{eq:Ham_new_block_renorm}).
Thus, we have presented five variants of the tSDRG algorithm.
In Sec.\ \ref{sec:numerics}, we precisely investigate numerical performance of tSDRGs depending on $\Delta$ for the Hamiltonian (\ref{Hamiltonian}) with Eqs. (\ref{eq_dist}) and (\ref{eq_dist_nonNN}).

\section{Numerical results}\label{sec:numerics}

In this section, we apply the tSDRG algorithms with different energy gaps proposed in Sec.\ \ref{sec:algorithm} to the random AF Heisenberg models on the 1D chain, square and triangular lattices.
We then compare their numerical performances and identify the optimal algorithm depending on parameter regimes of each lattice model.

\subsection{Details of calculations}\label{subsec:results_Method}

For the Hamiltonian (\ref{Hamiltonian}) with a random sample set of $J_{i,j}$ generated according to Eqs. (\ref{eq_dist}) and (\ref{eq_dist_nonNN}), we perform tSDRG calculations with the five variants of gaps, $\{ \Delta^{\! \rm I}_{\rm max}, \Delta^{\! \rm I}_{\rm gs}, \Delta^{\! \rm P}_{\rm max}, \Delta^{\! \rm P}_{\rm gs}, \Delta^{\! \rm P}_{\rm cut} \}$ .
The number of random samples is $\mathcal{N}_{\rm s} = 1000$.
The number of spins is $N=24$ for the 1D chain, $N=36$ ($6 \times 6$) for the square lattice, and $N=24$ (the shape is as in Ref.~\onlinecite{WuGS2019}) and $36$ ($6 \times 6$) for the triangular lattice.
The periodic boundary conditions are imposed for the all systems.
The number of kept states (the bond dimension in the tensor-network language) is up to $\chi = 80$.

After tSDRG computations for the ${\cal N}_s$ samples, we then take the random average of the following physical quantities to evaluate  accuracy of the algorithms.
First, let $E^{\, \nu}_\Delta$ be the ground-state energy of $\nu$th sample calculated by the tSDRG with a gap $\Delta \in \{ \Delta^{\! \rm I}_{\rm max}, \Delta^{\! \rm I}_{\rm gs}, \Delta^{\! \rm P}_{\rm max}, \Delta^{\! \rm P}_{\rm gs}, \Delta^{\! \rm P}_{\rm cut}  \}$.
We then consider the random average of the ground-state energy,
\begin{align}
  \overline{E}_\Delta \equiv \frac{1}{\mathcal{N}_{\rm s}}\sum_{\nu=1}^{\mathcal{N}_{\rm s}} E^{\,\nu}_\Delta \label{eq_E} \, .
\end{align}
For example, $\overline{E}_{\Delta^{\! \rm I}_{\rm max}}$ represents the random-averaged ground-state energy calculated with the tSDRG with $\Delta^{\! \rm I}_{\rm max}$. 
Second, we compute the random average of errors of the ground-state energies per spin,
\begin{align}
  \overline{\delta e}_\Delta \equiv 
  \frac{1}{\mathcal{N}_{\rm s} N }\sum_{\nu=1}^{\mathcal{N}_{\rm s}}
\left( E^{\, \nu}_\Delta - E_{\rm exact}^{\, \nu} \right), \label{eq_delta_E}
\end{align}
where $E_{\rm exact}^{\, \nu}$ is the exact ground-state energies obtained by the ED or QMC methods (see below).
Note that $\overline{\delta e}_\Delta \ge 0$, since $E_{\rm exact}^{\, \nu}$ is the trivial lower bound of $E^{\, \nu}_\Delta$.
Third, we calculate the random average of errors of the ground-state correlation functions,
\begin{align}
  \overline{\delta g}_\Delta \equiv
  \frac{1}{\mathcal{N}_{\rm s}}\sum_{\nu=1}^{\mathcal{N}_{\rm s}}
  \sqrt{\frac{2}{N(N-1)} \sum_{i \ne j}
    \left[g^{\, \nu}_\Delta(i,j) - g^{\nu}_{\rm exact}(i, j)\right]^2}, \label{eq_delta_G}
\end{align}
\begin{align}
  g^{\,\nu}_\Delta(i, j) &\equiv \left\langle \bm{S}_i \cdot \bm{S}_j \right\rangle^{\, \nu}_\Delta,\\
  g^{\,\nu}_{\rm exact}(i, j) &\equiv \left\langle \bm{S}_i \cdot \bm{S}_j \right\rangle_{\rm exact}^{\, \nu},
\end{align}
where $\langle\cdots\rangle^{\, \nu}$ ($\langle\cdots\rangle_{\rm exact}^{\, \nu}$)  denotes the ground-state expectation value of $\nu$th sample obtained by the tSDRG with $\Delta$ (ED or QMC).
Finally, for the square and triangular lattices, we calculate the random average of the static spin structure factor,
\begin {align}
  \overline{S}_{\Delta}(\bm{q})
  &\equiv
  \frac{1}{\mathcal{N}_{\rm s}}\sum_{\nu=1}^{\mathcal{N}_{\rm s}}
  \left\langle \left|
  \frac{1}{N}\sum_j {\bm{S}}_j e^{{\rm i} \bm{q} \cdot {\bm{R}}_j}
  \right|^2 \right\rangle_\nu \nonumber \\
  &= 
  \frac{1}{\mathcal{N}_{\rm s}N}\sum_{\nu=1}^{\mathcal{N}_{\rm s}}
  \sum_{i, j} g^{\, \nu}_\Delta(i, j)
  \cos \left[\bm{q} \cdot ({\bm{r}}_i-{\bm{r}}_j) \right].
\label{eq_Sq}
\end {align}

In order to calculate $E_{\rm exact}^{\, \nu}$ and  $ g_{\rm exact}^{\, \nu}(i, j) $ in  Eqs. (\ref{eq_delta_E}) and (\ref{eq_delta_G}), we use the ED method for the 1D chain and triangular lattice systems, and the loop-type QMC algorithm\cite{SyljuaasenS2002, KawashimaH2004} for the square lattice systems.
In QMC simulations,  the Monte Carlo (MC) average was taken over $10^5$ MC samples for every system with the given set of $J_{i,j}$, so that the MC errors are negligible in the scale of $\overline{\delta e}_\Delta$ and $\overline{\delta g}_\Delta$.
Moreover, the temperature used in the QMC simulations is $k_BT=1/128$, which is low enough to extract the ground-state properties of the finite size systems.
We have actually confirmed that the tSDRG results of the ground state energy are always lower bounded by the QMC results for the all random samples.

\subsection{the one-dimensional chain}\label{subsec:results_1D}

The bulk ground state of the random AF Heisenberg chain is qualitatively characterized by the random-singlet fixed point extracted by the perturbative SDRG\cite{MaDH1979, DasguptaM1980, Fisher1994}.
Moreover, the tSDRG with $\Delta^{\! \rm P}_{\rm cut}$ leads to a quantitative description of the ground state in the finite-size system level\cite{HikiharaFS1999}.
Comparing the tSDRG results with various $\Delta$, here, we demonstrate that the one with $\Delta^{\! \rm I}_{\rm max}$ further improves its numerical accuracy in the strong randomness regime.

\begin{figure}[tb]
  \centering\includegraphics[width=\linewidth]{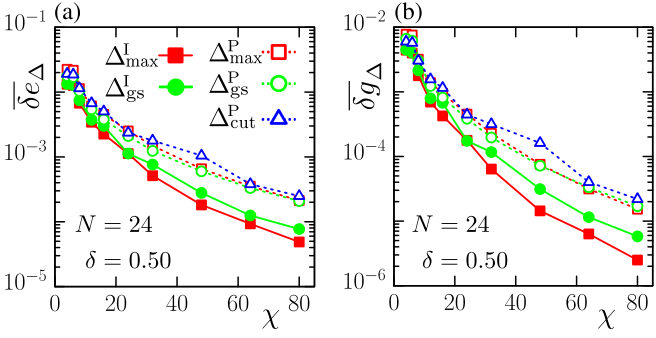}
  \caption{
    $\chi$-dependences of accuracy of the tSDRGs with various $\Delta$ for the random AF Heisenberg chain of $\delta=0.50$ and $N=24$: 
    (a) Error of the ground-state energy, $\overline{\delta e}_\Delta$, and (b) error of the ground-state correlation function, $\overline{\delta g}_\Delta$.
    Solid symbols represent the results of the tSDRGs with $\Delta^{\! \rm I}_{\rm max}$ and $\Delta^{\! \rm I}_{\rm gs}$, while open symbols indicate the results for $\Delta^{\! \rm P}_{\rm max}$, $\Delta^{\! \rm P}_{\rm gs}$, and $\Delta^{\! \rm P}_{\rm cut}$.
    Error bars due to the random average are negligible compared with the symbols.
  }\label{fig_1D}
\end{figure}

\begin{figure}[tb]
  \centering\includegraphics[width=\linewidth]{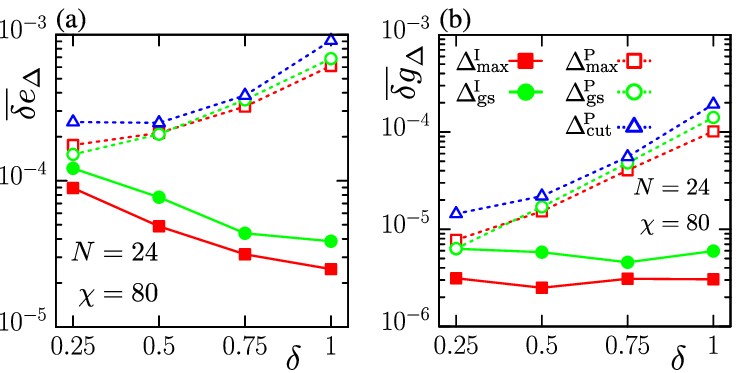}
  \caption{
 $\delta$-dependences of accuracy of the tSDRGs with various $\Delta$  at $\chi=80$ for the random AF Heisenberg chain of $N=24$: 
    (a) Error of the ground-state energy, $\overline{\delta e}_\Delta$, and (b) error of the ground-state correlation function, $\overline{\delta g}_\Delta$.
    Solid symbols represent the results of the tSDRGs with $\Delta^{\! \rm I}_{\rm max}$ and $\Delta^{\! \rm I}_{\rm gs}$, while open symbols indicate the results for $\Delta^{\! \rm P}_{\rm max}$, $\Delta^{\! \rm P}_{\rm gs}$, and $\Delta^{\! \rm P}_{\rm cut}$.
    Error bars are negligible compared with the symbols.
  }\label{fig_1D_delta}
\end{figure}

Figure \ref{fig_1D} shows the error of the ground-state energy $\overline{\delta e}_\Delta $, and the error of the correlation functions $\overline{\delta g}_\Delta$, for the $N=24$ chain with the randomness parameter $\delta = 0.5$.
In Fig. \ref{fig_1D_delta}, we also show the $\delta$-dependences of $\overline{\delta e}_\Delta$ and $\overline{\delta g}_\Delta$ at $\chi=80$.
Note that for the ${\cal N}_s=1000$ samples, the mean value of the exact energy is $\overline{E}_{\rm exact}= -11.04$ and its dispersion is $\sqrt{ \overline{(E_{\rm exact})^2} - (\overline{E}_{\rm exact})^2 }  = 0.42$.
We also note that the error bars due to the random average are not shown in Figs. \ref{fig_1D} and \ref{fig_1D_delta}, since they are sufficiently small (about $5\%$ of $\overline{\delta e}_{\Delta}$  or $\overline{\delta g}_{\Delta}$) for ${\cal N}_s=1000$.

In Fig. \ref{fig_1D}, we see that as $\chi$ increases, all of $\overline{\delta e}_\Delta$ and $\overline{\delta g}_\Delta$ rapidly approach zero, implying that the tSDRGs basically capture the correct ground state for the 1D case.
From the viewpoint of practical numerical computation, nevertheless, an important point is that the tSDRG algorithms with $\Delta^{\! \rm I}_{\rm max}$ and $\Delta^{\! \rm I}_{\rm gs}$ provide more accurate results, and this tendency becomes prominent in large $\delta$ regime in Fig. \ref{fig_1D_delta}. 
In particular, the algorithm with $\Delta^{\! \rm I}_{\rm max}$ achieves the best accuracy. 
At $\delta=1.0$ in Fig. \ref{fig_1D_delta}, for example,  the tSDRG algorithm with $\Delta^{\! \rm I}_{\rm max}$ yields $\overline{\delta e}_{\Delta^{\! \rm I}_{\rm \! max}} = 2.5(3) \times 10^{-5}$  and $\overline{\delta g}_{\Delta^{\! \rm I}_{\rm \!max}} = 3.1(5)\times 10^{-6}$, which are significantly improved from the values, $\overline{\delta e}_{\Delta^{\! \rm P}_{\rm \! cut}} = 9.1(7) \times 10^{-4}$  and $\overline{\delta g}_{\Delta^{\! \rm P}_{\rm \!  cut} } = 1.9(2) \times 10^{-4}$ obtained by the previous algorithm with $\Delta^{\! \rm P}_{\rm \!  cut}$.

In the context of tensor network, such an improvement of the accuracy suggests that essential information for the network structure representing the random singlet state is embedded in  $\mathcal{H}^{\rm I}$ rather than  $\mathcal{H}^{\rm P}$.
Moreover, we have confirmed that during tSDRG iterations with use of $\Delta^{\!  \rm I}_{\rm max}$ for $\delta=1.0$ and $\chi=80$,  about 95\% of $\Delta^{\! \rm I}_{\rm max}$ coincides with  $\Delta^{\! \rm I}_{\rm gs}$ in the spectra of ${\cal H}^{\rm I}$.
Indeed, the tSDRG with $\Delta^{\! \rm I}_{\rm gs}$ achieves accuracy close to but slightly worse than that with $\Delta^{\!  \rm I}_{\rm max}$.
This suggests that the gap structures in the higher energy spectra are also relevant for improving the numerical accuracy.

\subsection{Square lattice}\label{subsec:results_Squ}

We discuss efficiency of the tSDRG algorithms for the square-lattice random AF Heisenberg model, comparing tSDRG results for $N=36$ with the quasi-exact results obtained by QMC simulations. 
Note that the square-lattice model exhibits the AF long-range order in the bulk limit even under strong randomness, although its magnitude is significantly reduced.\cite{LaflorencieWLR2006}

\begin{figure}[tb]
  \centering\includegraphics[width=\linewidth]{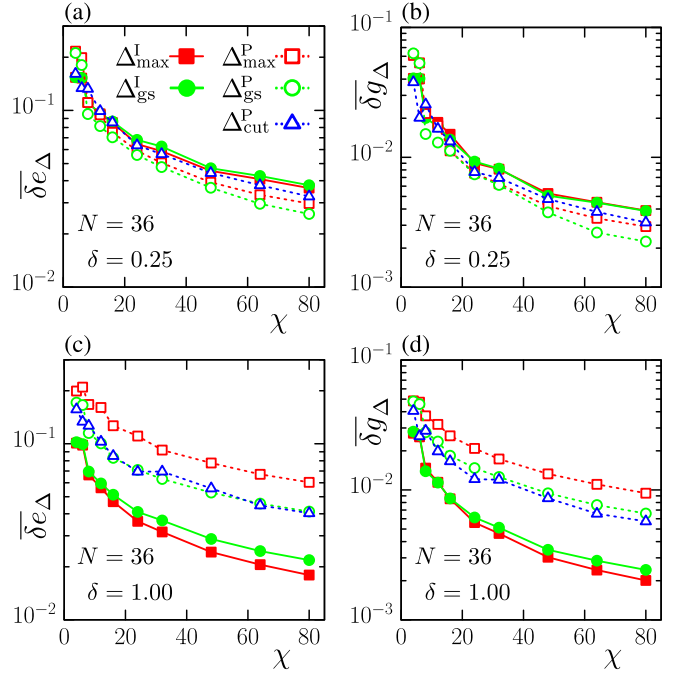}
  \caption{
    $\chi$-dependences of accuracy of the tSDRG algorithms with various $\Delta$ for the  random AF Heisenberg model on the $N=36$ square lattice:
    (a) Error of the ground-state energy, $\overline{\delta e}_{\rm \Delta}$, and (b) error of the ground-state correlation function, $\overline{\delta g}_{\rm \Delta}$, for $\delta=0.25$.
    Panels (c) and (d) respectively represent  $\overline{\delta e}_{\rm \Delta}$ and  $\overline{\delta g}_{\rm \Delta}$ for $\delta=1.00$.
    Error bars are negligible compared with the symbols.
  }\label{fig_square_error}
\end{figure}

In Fig. \ref{fig_square_error}, we show the tSDRG results for the $N=36$ systems of two typical values of $\delta = 0.25$ and  $1.00$,  which  respectively correspond to weak and strong randomness regimes.
In the both cases, the errors $\overline{\delta e}_\Delta$ and ${\overline{\delta g}}_\Delta$ gradually decrease as $\chi$ increases.
In contrast to the 1D case, however, ranking of the accuracy for different $\Delta$ depends on $\delta$; 
In the weak randomness case ($\delta=0.25$), the tSDRG algorithm with $\Delta^{\! \rm P}_{\rm gs}$ yields the most accurate result, while in the strong randomness case ($\delta=1.00$), the results with $\Delta^{\! \rm I}_{\rm max}$ turn out to be the most accurate.
For example, the ratios of the errors at $\chi=80$ are 
\[ \frac{\overline{\delta e}_{ \Delta^{\! \rm P}_{\rm \! gs} } }{ \overline{\delta e}_{ \Delta^{\! \rm I}_{\rm \! max} }  }\simeq 71\% 
\quad {\rm and }\quad  
\frac{  \overline{\delta g}_{\Delta^{\! \rm P}_{\rm \! gs}} }{  \overline{\delta g}_{ \Delta^{\! \rm I}_{\rm \! max} } }\simeq 58\% \, ,
\]
for $\delta = 0.25$. 
By contrast, for $\delta=1.0$, we have 
\[ \frac{ \overline{\delta e}_{ \Delta^{\! \rm I}_{\rm \! max} } }{ \overline{\delta e}_{ \Delta^{\! \rm P}_{\rm \! gs} } }\simeq 44\% 
\quad {\rm and }\quad  
\frac{ \overline{\delta g}_{ \Delta^{\! \rm I}_{\rm \! max} } }{ \overline{\delta g}_{ \Delta^{\! \rm P}_{\rm \! gs} } }\simeq 31\% \, ,
\]
which implies that the improvement of the tSDRG based on $\mathcal{H}^{\rm I}$ becomes more prominent in the strong randomness regime.

\begin{figure}[tb]
  \centering\includegraphics[width=\linewidth]{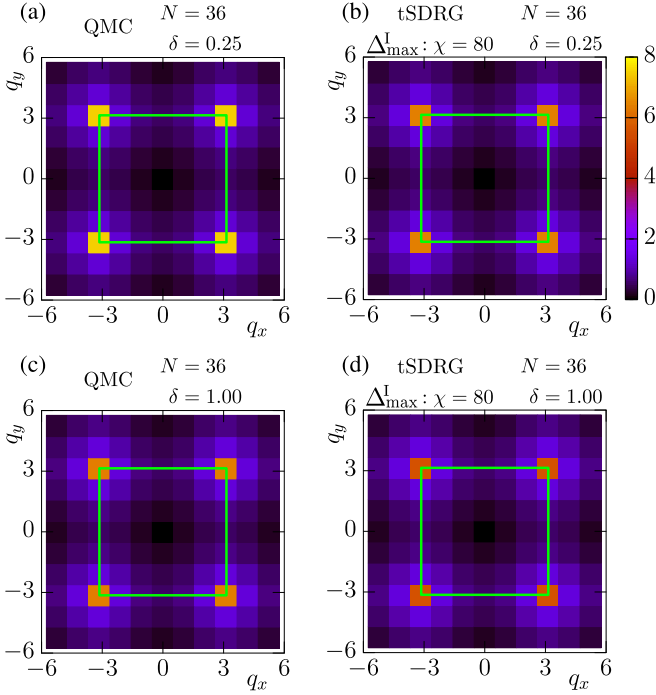}
  \caption{
    Intensity plots of the static spin structure factor $\overline{S}(\bm q)$ for the square-lattice Heisenberg model of $N=36$:
    (a) QMC  result and  (b) tSDRG result with $\Delta^{\! \rm I}_{\rm max}$ of $\chi=80$, for $\delta=0.25$.
    Panels (c) and (d) respectively show QMC result and tSDRG result of $\chi=80$ with $\Delta^{\! \rm I}_{\rm max}$ for $\delta=1.00$.
    The green line shows the boundary of the first Brillouin zone of the square lattice.
  }\label{fig_square_Sq}
\end{figure}

In order to illustrate the qualitative features of the tSDRG results, we present comparisons of $\overline{S}_{\rm \Delta^{\! \rm I}_{\rm max}} (\bm{q})$ with the exact $\overline{S}(\bm{q})$ calculated by the QMC in Fig. \ref{fig_square_Sq}.
For both of $\delta = 0.25$ and $1.00$, we can verify that the tSDRG results basically capture the correct ground-state properties such as the peak structures at $\bm{q} = (\pi, \pi)$.
However, the reduction of the peak height of $\overline{S}_{\Delta^{\! \rm I}_{\rm max} }(\bm{q})$ is slightly large at  $\delta=0.25$ where the tSDRG with $\Delta^{\! \rm I}_{\rm gs}$ is more accurate in the weak randomness regime.

\begin{figure}[tb]
  \centering\includegraphics[width=0.7\linewidth]{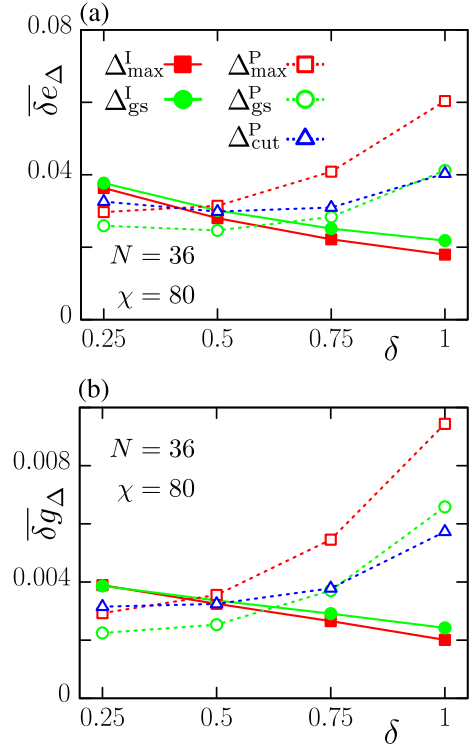}
  \caption{
 $\delta$-dependences of accuracy of the tSDRGs with various $\Delta$  at $\chi=80$ for the random AF Heisenberg model on the $N=36$ square lattice:
    (a) Error of the ground-state energy, $\overline{\delta e}_\Delta$, and (b) error of the ground-state correlation function, $\overline{\delta g}_\Delta$.
    Solid symbols represent the tSDRG results with $\Delta^{\! \rm I}_{\rm max}$ and $\Delta^{\! \rm I}_{\rm gs}$, while open symbols indicate the results for $\Delta^{\! \rm P}_{\rm max}$, $\Delta^{\! \rm P}_{\rm gs}$, and $\Delta^{\! \rm P}_{\rm cut}$.
    Error bars are negligible compared with the symbols.
  }\label{fig_square_delta}
\end{figure}

In Fig.\ \ref{fig_square_delta}, we summarize the $\delta$ dependences of $\overline{\delta e}_\Delta$ and $\overline{\delta g}_\Delta$.
In the weak randomness regime ($\delta \lesssim 0.6$), the tSDRG algorithms based on  $\mathcal{H}^{\rm P}$ exhibit slightly better accuracy than the ones based on $\mathcal{H}^{\rm I}$. 
As $\delta$ increases, however, the accuracy of the tSDRG algorithms based on the spectrum of $\mathcal{H}^{\rm I}$ is monotonously improved.
In particular, the algorithm with $\Delta^{\! \rm I}_{\rm max}$ exhibits the best accuracy for $\delta \gtrsim  0.6$ among various $\Delta$, as in the case of the 1D chain.
This suggests that the gap $\Delta^{\! \rm I}_{\rm max}$ in the spectrum of $\mathcal{H}^{\rm I}$ provides a tree-tensor network structure suitable for the ground state in the strong randomness, while the gap $\Delta^{\! \rm P}_{\rm gs}$ for $\mathcal{H}^{\rm P}$ may rather efficient for representing the short range N\'{e}el order.
However, we remark that the above analyses are based on the results for $N=36$ systems, which are insufficient to discuss the bulk fixed point of the tSDRG algorithms.
A further analysis of the fixed point properties should be an important future issue.

\subsection{Triangular lattice}\label{subsec:results_Tri}

We analyze the numerical accuracy of the tSDRG algorithms for the triangular-lattice random AF Heisenberg model.
For the triangular-lattice model, a quantum phase transition between the 120$^\circ$ magnetic ordered phase and the randomness-induced disordered phase was suggested around $\delta \sim 0.5$, on the basis of ED calculations up to $N=24$.\cite{WatanabeKNS2014, ShimokawaWK2015, WuGS2019}

\begin{figure}[tb]
  \centering\includegraphics[width=\linewidth]{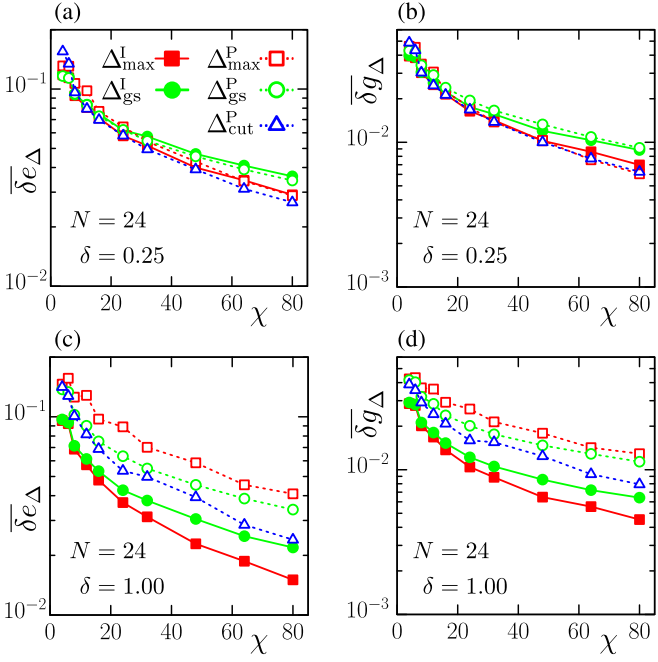}
  \caption{
    $\chi$-dependences of accuracy of the tSDRG algorithms with various $\Delta$ for the random AF Heisenberg model on the $N=24$ triangular lattice: 
    (a) Error of the ground-state energy, $\overline{\delta e}_{\rm \Delta}$, and (b) error of the ground-state correlation function, $\overline{\delta g}_{\rm \Delta}$, for $\delta=0.25$.
    Panels (c) and (d) respectively represent  $\overline{\delta e}_{\rm \Delta}$ and  $\overline{\delta g}_{\rm \Delta}$ for $\delta=1.00$.
    Error bars are negligible compared with the symbols.
  }\label{fig_triangular_error}
\end{figure}

Figure \ref{fig_triangular_error} shows the results of $\overline{\delta e}_\Delta$ and $\overline{\delta g}_\Delta$ for the $N=24$ systems of $\delta = 0.25$ and  $1.00$, where both of $\overline{\delta e}_\Delta$ and $\overline{\delta g}_\Delta$  gradually decrease with increasing $\chi$.
In the weak randomness phase ($\delta =0.25$), the accuracy of tSDRG with $\Delta^{\! \rm P}_{\rm cut}$ is slightly better than the others, though there is no significant difference among all $\Delta$.
By contrast, for the strong randomness ($\delta =1.00$), the results of $\Delta^{\! \rm I}_{\rm max}$ are clearly better than the others; the ratios of the errors at $\chi= 80$ for $\delta=1.00$ are given by 
\[ \frac{\overline{\delta e}_{\Delta^{\! \rm I}_{\rm max}} }{  \overline{\delta e}_{\Delta^{\! \rm P}_{\rm cut}} } \simeq 63\% \, ,\quad {\rm and} \quad
\frac{ \overline{\delta g}_{\Delta^{\! \rm I}_{\rm max}} }{  \overline{\delta g}_{\Delta^{\! \rm P}_{\rm cut}}  } \simeq 57\% \, ,
\]
which demonstrate a clear advantage of the tSDRG with $\Delta^{\! \rm I}_{\rm max}$ in the strong randomness regime.

\begin{figure}[tb]
  \centering\includegraphics[width=\linewidth]{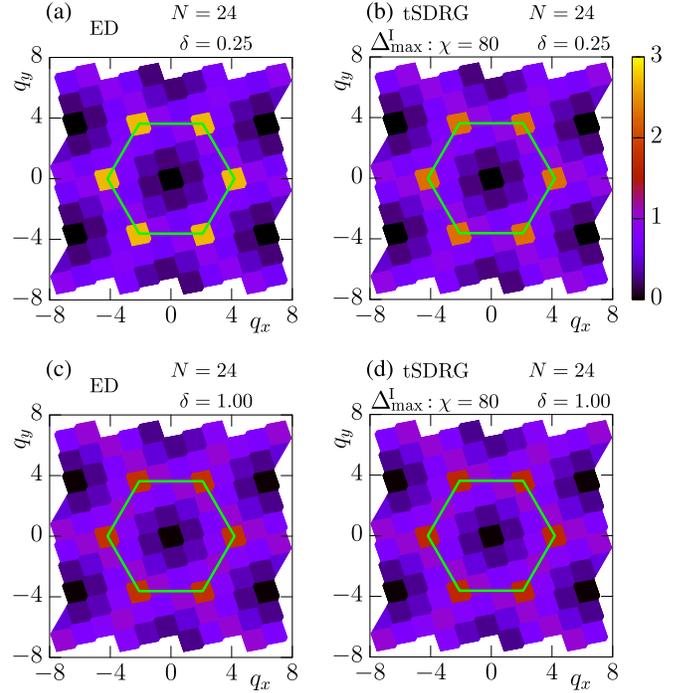}
  \caption{
    Intensity plots of the static spin structure factor $\overline{S}(\bm q)$ for the $N=24$ triangular-lattice Heisenberg model:
    (a) ED result and (b) tSDRG result of $\chi=80$ with $\Delta^{\! \rm I}_{\rm max}$ for $\delta=0.25$.
    Panels (c) and (d) respectively show ED result and tSDRG result of  $\chi = 80$ with $\Delta^{\! \rm I}_{\rm max}$ for $\delta=1.00$.
    The green line shows the boundary of the first Brillouin zone of the triangular lattice.
  }\label{fig_triangular_Sq}
\end{figure}

In order to see the qualitative features of the tSDRG results, we present the comparison of $S_{\Delta^{\! \rm I}_{\rm max} }(\bm{q})$ at $\chi=80$ with ED results in Fig. \ref{fig_triangular_Sq}.
For $\delta=0.25$, $S_{\Delta^{\! \rm I}_{\rm max} }(\bm{q})$ exhibits clear peaks at K points, which is consistent with the ED result that the system is in the 120$^\circ$ ordered phase.
For $\delta=1.00$, meanwhile, the peaks at K points are significantly reduced by the strong randomness, which suggests that the randomness-induced disordered phase may be realized in the bulk limit.
We thus think that the tSDRG with $\Delta^{\! \rm I}_{\rm max}$ successfully reproduces the qualitative features of the static spin structure factor for the triangular lattice model.

\begin{figure}[tb]
  \centering\includegraphics[width=0.7\linewidth]{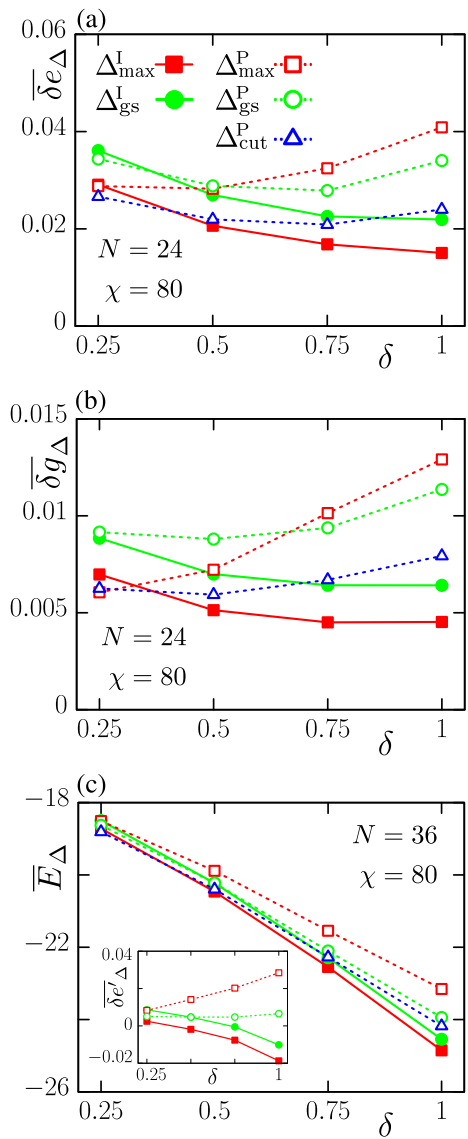}
  \caption{
$\delta$-dependences of accuracy of the tSDRGs with various $\Delta$ at $\chi=80$ for the random AF Heisenberg model on triangular lattices:
    (a) Error of the ground-state energy, $\overline{\delta e}_\Delta$, and (b) error of the ground-state correlation function, $\overline{\delta g}_\Delta$, for $N=24$.
    Solid symbols represent the results of the tSDRGs with $\Delta^{\! \rm I}_{\rm max}$ and $\Delta^{\! \rm I}_{\rm gs}$, while open symbols indicate the results for $\Delta^{\! \rm P}_{\rm max}$, $\Delta^{\! \rm P}_{\rm gs}$, and $\Delta^{\! \rm P}_{\rm cut}$.
    Panel (c) shows $\delta$-dependences of the ground-state energy $\overline{E}_\Delta$ for $N=36$ and $\chi = 80$.
    The inset shows the $\delta$-dependence of $\overline{\delta e'}_\Delta$, which is the energy gain from $\Delta^{\! \rm P}_{\rm cut}$ defined by Eq. (\ref{delta_e'}).
    Error bars are negligible compared with the symbols.
  }\label{fig_triangular_delta}
\end{figure}

In Figs. \ref{fig_triangular_delta}(a) and (b), we summarize the $\delta$-dependences of $\overline{\delta e}_\Delta$ and $\overline{\delta g}_\Delta$ at $\chi=80$, which exhibit basically the same tendency as those for the square lattice.
The tSDRG with $\Delta^{\! \rm P}_{\rm cut}$ achieves accuracy slightly better than the others in the weak randomness regime, where the magnetic order was suggested in Refs. \onlinecite{WatanabeKNS2014, ShimokawaWK2015, WuGS2019}, while the tSDRG algorithm with $\Delta^{\! \rm I}_{\rm max}$  becomes the most accurate in the strong randomness regime ($\delta \gtrsim 0.4$).
We have confirmed that, as in the case of the 1D chain, about 88\% of $\Delta^{\! \rm I}_{\rm max}$ coincides with $\Delta^{\! \rm I}_{\rm gs}$ during tSDRG iterations with $\Delta^{\! \rm I}_{\rm max}$. 
Thus, some nontrivial structures embeded in higher excitations can be properly built in the tSDRG computation with $\Delta^{\! \rm I}_{\rm max}$ compared to the one with $\Delta^{\! \rm I}_{\rm gs}$, resulting in the accuracy for $\Delta^{\! \rm I}_{\rm max}$ slightly better than that for $\Delta^{\! \rm I}_{\rm gs}$.

In Fig.\ \ref{fig_triangular_delta}(c), we present the $\delta$ dependences of the ground-state energies $\overline{E}_\Delta$ for $N=36$, which is the system size inaccessible by the ED. 
In the inset, we also show the random average of the energy gain from the previous version of tSDRG with $\Delta^{\! \rm P}_{\rm cut}$, which is defined as
\begin{align}
  \overline{\delta e'}_\Delta \equiv 
  \frac{1}{\mathcal{N}_{\rm s} N}\sum_{\nu=1}^{\mathcal{N}_{\rm s}}
  \left( E^{\, \nu}_{\Delta} - E_{\Delta^{\! \rm P}_{\rm cut}}^{\, \nu} \right) .
\label{delta_e'}
\end{align}
We find that the order of the accuracy for various $\Delta$ is consistent with those of $\overline{\delta e}_\Delta$ for $N=24$. 
Thus, we can conclude again that the tSDRG algorithm with $\Delta^{\! \rm I}_{\rm max}$ extracted from $\mathcal{H}^{\rm I}$ is efficient in the strong randomness regime.
However, it should be noted that a further investigation of the tensor network structures is clearly required for directly determining the phase boundary between the 120$^\circ$ ordered phase and the disordered phase in the context of the fixed point of the tSDRG algorithms.

\section{Summary and discussions}\label{sec:Summary}

For the purpose of formulating an efficient numerical method for 2D random spin systems, we have systematically discussed a possible improvement of the tSDRG algorithm.
We have particularly focused on the determination process of the block pair to be merged, which is relevant to the network structure of tensors crucial for the resulting accuracy.
We have proposed five variants of tSDRG algorithm, in which the strength of links connecting blocks are quantified by energy gaps $\Delta \in \{ \Delta^{\! \rm I}_{\rm max}, \Delta^{\! \rm I}_{\rm gs}, \Delta^{\! \rm P}_{\rm max}, \Delta^{\! \rm P}_{\rm gs}, \Delta^{\! \rm P}_{\rm cut} \} $ embedded in the spectra of the interaction Hamiltonian (${\cal H}^{\rm I}$) or in the pair bock Hamiltonian (${\cal H}^{\rm P}$).
We have then examined their numerical accuracy for the random AF Heisenberg models on 1D and 2D lattices.
For the 1D chain, we have confirmed that the tSDRG algorithm with $\Delta^{\! \rm I}_{\rm max}$ further improves its accuracy, compared with the previous tSDRG algorithm with  $\Delta^{\! \rm P}_{\rm cut}$ that was already known to be efficient in 1D random spin models\cite{HikiharaFS1999,GoldsbroughR2014}.
For both of square and triangular lattices,  by comparing the tSDRG results to the exact data obtained with the exact diagonalization and quantum Monte Carlo methods, we have also demonstrated that the algorithm with $\Delta^{\! \rm I}_{\rm max}$ provides prominently accurate results in the strong randomness regime, while for the weak randomness case, the accuracy of the previous algorithm with $\Delta^{\! \rm P}_{\rm cut}$ is almost the same as or better than that with $\Delta^{\! \rm I}_{\rm max}$.
In practical sense, thus, it is concluded that we should use the tSDRG algorithms with an appropriate gap $\Delta$, depending on the amplitude of randomness.

From the viewpoint of renormalization group, an interesting implication of the present analysis is that the tSDRG algorithms based on ${\cal H}^{\rm I}$ yield fairly better results than those based on ${\cal H}^{\rm P}$ in the strong randomness regime, independently of the lattice structures. 
This suggests that the tSDRG with ${\cal H}^{\rm I}$ could generate tree-tensor networks suitable for describing such a randomness-induced disordered state  as random singlet state, where the entanglements among local singlet pairs are decoupled from each other.
In the weak randomness regime where the classical spin orders are rather robust, on the other hand, the tSDRG based on ${\cal H}^{\rm P}$ works slightly better. 
This may be because $\Delta^{\! \rm I}_{\rm max}$ extracted from ${\cal H}^{\rm I}$ is more likely to capture  such an order defined on the link  as spin-singlet state, while ${\cal H}^{\rm P}$ is likely to involve information of the onsite block state relevant to describing the classical local magnetic moment.
We should however note that the arguments above cannot directly refer to the bulk fixed point structure of the present tSDRG particularly in two dimensions.
In the tree-tensor network, a low-level branch of tensors is disconnected from the main tree network by cutting a single bond of a finite bond dimension $\chi$, implying that the entanglement entropy between the branch and the rest of the tree capable in the tSDRG algorithm is always bounded by $\sim \ln\ \chi$.
This is actually the case at the random-singlet fixed point of the 1D chain, where the tSDRG achieves very good accuracy. 
However, entanglement structures of bulk ground states of the 2D models under the strong randomness are still a nontrivial problem.
We need further researches to clarify whether or not the fixed point structure generated within the framework of the present tSDRG algorithm is appropriate for bulk 2D systems.

Finally, we comment on further improvements of the tSDRG algorithm.
A straightforward improvement is to iteratively optimize the renormalization matrix $V$ (isometry in the context of tensor network) in the tree-tensor network until the variational energy converges.\cite{TagliacozzoEV2009}
Another important approach is to directly refer to entanglements between the blocks in finding out the block pair to be merged.
Although the full treatment of the entanglement is a hard problem in a practical situation where the ground-state wavefunction of the whole system is not known a priori, we can formulate an entanglement-based tSDRG algorithm taking account of entanglements between neighboring blocks.
The details will be published elsewhere.\cite{SekiHO2020}
We also note that there is a proposal to apply other type of tensor network such as disordered multiscale entanglement renormalization ansatz to the random spin systems.\cite{GoldsboroughE2017}
This method can achieve more precise calculation than the tSDRG, in principle, but its computational cost is also much larger than tSDRG especially for 2D systems.

\begin{acknowledgments}
We would like to thank Tomotoshi Nishino and Hiroshi Ueda for fruitful discussions.
This work was supported by JSPS KAKENHI Grant Numbers 15K05198, 17H02931, and 19K03664.
\end{acknowledgments}


\begin{thebibliography}{37}
\bibitem [{not({\natexlab{b}})}]{note_early_stage}%
  \BibitemOpen
  \bibinfo {note} {In the early stage of tSDRG iterations, the dimension of the
  Hilbert space of the block pair to be merged is often less than $\chi$. In
  that case, we keep all the eigenstates of $\mathcal{H}^{\rm P}_{R+R'}$ in the
  renormalization process.}\BibitemShut {Stop}%
\bibitem [{not({\natexlab{c}})}]{note_Delta_cut}%
  \BibitemOpen
  \bibinfo {note} {A particular exception handling in the tSDRG with
  ${\Delta}^{\! \rm P}_{\rm cut}$ is that we cannot often define $\Delta^{\!
  \rm P}_{\rm cut}$ in the early stage of iterations, since the Hirbert space
  dimension of block pair is less than or equal to $\chi$, If this is the case,
  we employ ${\Delta}^{\! \rm P}_{\rm gs}$ instead of ${\Delta}^{\! \rm P}_{\rm
  cut}$.}\BibitemShut {Stop}%
\end{thebibliography}
\end{document}